\documentclass[aps,prl,preprint,groupedaddress]{revtex4-1}

\begin{document}


\title{Comments on the compositional boundary condition for diffuse interface model of a contact line}


\author{Lionel Hirschberg and Avraham Hirschberg}
\email[]{l.hirschberg@me.com}
\affiliation{Technische Universiteit Eindhoven}


\date{\today}

\begin{abstract}
The diffuse interface model of Cahn-Hilliard-van der Waals is often used to study various aspects of multi-phase flows such as droplets coalescence and contact line dynamics. The original model of Cahn-Hilliard-van der Waals uses an approximation which neglects some surface contributions to the free energy of the system and is justified by the large distance between the region of observation and the outer surface of the system. This is not a priori accurate when considering the dynamics of the contact line on a solid surface. A modification of the so called natural compositional boundary conditions obtained by minimization of the surface contribution to the free energy of the system is proposed for the vapor-liquid interface of a single component and for a quasi-incompressible binary mixture of partially miscible liquids. The results are obtained for a stagnant fluid in thermodynamic equilibrium but should be valid in a flow if local thermodynamic equilibrium can be assumed.
\end{abstract}


\maketitle

\section{Introduction}
\label{Introduction}
A sharp interface continuum model cannot describe the movement of a contact line on a solid surface when the so called no-slip boundary condition is imposed at the wall. Various modifications of the sharp interface model have been  described in review papers (\citep{Bonn}, \citep{Yisui}, \citep{Snoeijer}). Seppecher \cite{Seppecher} and Jacqmin \cite{Jacqmin} demonstrated that a diffuse interface model based on the modification of the theory of van der Waals \cite{vdWaals} by Cahn and Hilliard \cite{Cahn_Hilliard}  does predict a motion of a contact line when a no-slip boundary condition is applied at the wall. The diffuse interface model uses a local thermodynamic equilibrium assumption in which the specific free energy of the fluid depends not only on the local thermodynamic properties of the fluid but also on the spatial gradients in these quantities. In the Cahn-Hiliard-van der Waals approximation the gradient length $L_g$ is assumed to be large compared to the molecular attraction length scale $L_m$. When considering isotropic fluids the first gradient term vanishes in a Taylor expansion of the specific free energy around its value for a uniform fluid. The first non-vanishing  non-local terms are a term proportional to the Laplacian of the fluid properties and a term proportional to the square of the gradient. When considering the total free energy of the system, the Laplacian term is split by partial integration into a bulk term proportional to the square of the gradient and a surface term. This surface term is neglected (\citep{vdWaals},\citep{Cahn_Hilliard}) using the argument that the region of interest is far removed from walls. Obviously this is questionable when considering contact line dynamics. The application of the diffuse interface theory to wetting in the seminal paper of Cahn \cite{Cahn} uses this approximation as do the more recent studies (\citep{Seppecher}, \citep{Jacqmin}, \citep{Qian}, \citep{Khatavkar}, \citep{Yue}, \cite{:/content/aip/journal/pof2/23/1/10.1063/1.3541806}, \citep{Carlson}, \citep{Lee}, \citep{Magaletti}, \citep{Sibley}). To the authors knowledge this problem has not yet been discussed in the literature.  

An advantage of the gradient theory is that in principle one could use a single set of equations through a two phase system. However as explained by Yue et al. \cite{Yue} in practice the spatial discretisation cannot be sufficient to resolve typical length scales in interfaces. Hence in principle only results independent from the exact value of the gradient length are physically relevant. While Yue et al. \cite{Yue} propose a method to compensate for errors due to too large interface thickness, the multi-scale approach of  Seppecher \cite{Seppecher} is promising as well. He matches an inner diffuse interface model to a far-field sharp interface model. Such a multi-scale model could be further extended to include local molecular phenomena near the contact line. An example of this is including a disjoining potential term making the interaction energy between the fluid and the solid dependent on the local thickness of
the liquid film. Also one could introduce stochastic source terms in the equations as proposed by Snoeijer and Andreotti \cite{Snoeijer}. We however limit our discussion to the most commonly used model proposed by Cahn \cite{Cahn} in which the surface free energy is a function of the local bulk fluid properties adjacent to the wall. Hence the model remains essentially a continuum model. We will show that in the case of Cahn's \cite{Cahn} model for a semi-infinite two phase fluid in contact with a wall the extra surface free energy term cannot be neglected.\\

As proposed by Anderson et al.  \cite{Anderson}  we limit our discussion to a stagnant fluid at thermodynamic equilibrium. The results obtained can then be used to derive equations of motion if one assumes local thermodynamic equilibrium \cite{Anderson,Mauri}.\\

 Firstly, the Cahn-Hilliard-van der Waals theory applied to the interface between a vapor and a liquid of a single component system will be discussed with focus on modified wall compositional boundary conditions. Secondly, results for the commonly studied quasi-incompressible binary mixture of partially miscible fluids \citep{Lowengrub} are summarized. Finally the incompressible binary regular solution model of Cahn \cite{Cahn} is discussed.

\section{vapor liquid interface}
As stated before we limit our discussion to a stagnant fluid  in thermodynamic equilibrium. The condition for  thermodynamic equilibrium for a single component isolated system of volume $V$ enclosed by a surface $\partial V$ is found by van der Waals \citep{vdWaals} by seeking for a maximum of entropy  at constant mass $M$ and internal energy $U$. Ignoring surface effects, this corresponds to a maximum of the entropy:
\begin{equation}
S=\int_V \rho s d^3x
\end{equation}
where $s$ is the specific entropy and $\rho$ is the fluid density, with the constrains:
\begin{equation}
M=\int_V \rho d^3x={\rm constant}
\end{equation}
and
\begin{equation}
U=\int_V \rho u d^3x={\rm constant}
\end{equation}
where  $u$ the specific internal energy.
The constrained maximum of the entropy $S$ can be written using the Lagrange multipliers method as:
\begin{equation}
\delta\int_V \rho\left(\frac{-\mu}{T}+\frac{u}{T} -s\right)d^3x=\frac{1}{T}\delta\int_V \rho (f-\mu) d^3x=0
\end{equation}
where  $f\equiv u-Ts$ is the specific Helmholtz free energy. By definition the Lagrangian multipliers $1/T$ and $\mu/T$ are constants. The Lagrangian multiplier $1/T$ is the inverse of the temperature, which in equilibrium is uniform within $V$. The Lagrange multiplier $\mu/T$ is the chemical potential  $\mu$ divided by the temperature, which in equilibrium is uniform over the system. 

We consider conditions such that two uniform phases phases coexist: a liquid phase of density $\rho_\alpha$ and its vapor of density $\rho_\beta$. The densities in the range $\rho_\beta<\rho<\rho_\alpha$  (between the spinodal points) are for uniform systems either unstable or meta-stable. 
The key idea of van der Waals \citep{vdWaals} is that in the interface of finite thickness between the liquid and the vapor all the densities in the range $\rho_\beta<\rho<\rho_\alpha$ are present and stable. Thus the non-uniformity of the fluid in the interface stabilizes the fluid, allowing static equilibrium. Hence, for the closed system considered, one should minimise the total free energy of the system taking into account the fact that the specific free energy of the fluid depends  not only on the local density but also on the gradients of the density: $f=f(\rho,{\nabla}\rho,{\nabla}{\nabla}\rho,\dots)$. The next step is to assume that the contribution of the gradients is small, in other words that the molecular length scale $L_m$ is small compared to the gradient length $L_g$ (the interface thickness) i.e.  $\epsilon\equiv L_m/L_g<<1$. For an isotropic fluid the leading order terms in a Taylor expansion of $f$ around a uniform state of density $\rho$ are given by:
\begin{equation}\label{Taylor}
f=f_0(\rho)+\left(\frac{\partial f}{\partial{\nabla}^2\rho}\right)_0{\nabla}^2\rho+\frac{1}{2}\left(\frac{\partial^2f}{\partial |{\nabla}\rho |^2}\right)_0|{\nabla}\rho |^2+O(\epsilon^3).
\end{equation}
The index $0$ indicates that the coefficients are functions of $\rho$ only.  As proposed by Cahn \cite{Cahn}, the interaction of the fluid with the wall surface $\partial V$ is assumed to be described by a free energy per unit surface $\hat f_w(\rho)$, which is a function of the density $\rho$ of the fluid at the wall (in a material element, large compared to $L_m$, adjacent to the wall). The molecular interaction between the fluid and the wall is assumed to be limited to a length scale $L_m$ (monolayer), short compared to other length scales in the problem \citep{Cahn}.  This leads to:
\begin{eqnarray}
&\delta &\int_V\rho\left\{f_0+\left[-\frac{1}{\rho}\frac{d}{d\rho}\left(\rho\left(\frac{\partial f}{\partial {\nabla}^2\rho}\right)_0\right)+\frac{1}{2}\left(\frac{\partial^2f}{\partial|{\nabla}\rho |^2}\right)\right]|{\nabla}\rho|^2-\mu\right\}d^3x~~\\ \nonumber
+&\delta&\int_{\partial V}\left[ f_w+\rho\left(\frac{\partial f}{\partial{\nabla}^2\rho}\right)_0({\nabla}\rho \cdot \vec n)\right]d^2x=0
\label{eq:Free_energy}
\end{eqnarray}
where $\vec{n}$ it the outer normal on $\partial V$. In this expression we have carried out a partial integration to express the term with the Laplacian $ {\nabla}^2\rho$ in the bulk of the fluid  in terms proportional to $|{\nabla} \rho |^2$ and a surface contribution proportional to $({\nabla}\rho \cdot \vec n)$. We neglected the mass of the fluid monolayer on the wall surface $\partial V$ compared to the mass of the fluid in $V$.

Using the fact that in a stagnant fluid $\delta ({\nabla}\rho)={\nabla}(\delta\rho)$, by carrying a second partial integration and considering that the perturbations $\delta\rho$ are arbitrary, we obtain:
\begin{equation}
\mu=\frac{d\rho f_0}{d \rho}+\frac{d \rho K}{d\rho}|{\nabla} \rho |^2-{\nabla}\cdot(\rho K{\nabla} \rho)
\end{equation}
with $K$ defined as is usual in the literature:
\begin{equation}
K=-\frac{2}{\rho}\frac{d}{d\rho}\left[\rho\left(\frac{\partial f}{\partial{\nabla}^2\rho}\right)_0\right]+\left(\frac{\partial^2f}{\partial|{\nabla}\rho |^2}\right)_0.
\end{equation} 
In the literature one often assumes that $K$ is a constant but in general it is a function of $\rho$.
After the second partial integration and application of the variation the surface integral becomes:

\begin{equation}
   \int_{\partial V}\left\{\left[\frac{d f_w}{d\rho}+\rho K ({\nabla}\rho \cdot \vec n)+\left(\frac{d}{d\rho}\rho\left(\frac{\partial f}{\partial {\nabla}^2\rho}\right)\right)({\nabla}\rho \cdot \vec n)\right]\delta\rho+\rho\left(\frac{\partial f}{\partial {\nabla}^2\rho}\right)_0\delta({\nabla}\rho \cdot \vec n)\right\}d^2x=0
   \end{equation}
Assuming that $({\nabla}\rho \cdot \vec n)$ is a function of $\rho$ and  consequently using $\delta ({\nabla}\rho \cdot \vec n)=[d(({\nabla}\rho \cdot \vec n)/d\rho]\delta\rho$ one obtains the compositional natural boundary condition in the form of a differental equation for $({\nabla}\rho \cdot \vec n)$:
\begin{equation}\label{vapor_liquid}
\frac{d f_w}{d\rho}+\rho K ({\nabla}\rho \cdot \vec n)+\left(\frac{d}{d\rho}\rho\left(\frac{\partial f}{\partial {\nabla}^2\rho}\right)\right)({\nabla}\rho \cdot \vec n)+\rho\left(\frac{\partial f}{\partial {\nabla}^2\rho}\right)_0\frac{d({\nabla}\rho \cdot \vec n)}{d\rho}=0.
\end{equation}
We should now provide an initial condition for the integration of this equation. Note that this integration is carried out in the thermodynamic space and can be carried out independently of any flow simulations. This integration determines $({\nabla}\rho \cdot \vec n)$ as a function of $\rho$. 
For a wetting wall ($d f_w/d\rho<0$) we expect that at very high densities the wall will be saturated. Let $\rho_{max}$ be the maximum thermodynamically allowable liquid phase density  at the temperature considered (limit of very high pressures).  In principle $\rho_{max}$ is outside the range of $\rho$ that will be found in the actual flow. We expect that  the derivative of $f_w$ with respect to $\rho$ vanishes  for $\rho=\rho_{max}$ (i.e. $(d f_w/d\rho)_{\rho_{max}}=0$) and 
that $({\nabla}\rho{\cdot \vec{n}})_{\rho_{max}}=0$. This ad hoc assumption should in principle be confirimed on the basis of physical models for the interaction of the fluid with the wall. Such an assumption could, however, also be imposed for convenience (without justification).
 
Note that Mauri \citep{Mauri} obtained a similar result but stated that the perturbation $\delta ({\nabla}\rho \cdot \vec n)$ can be assumed to vanish. This results into an explicit equation for $({\nabla}\rho \cdot \vec n)$ as a function of $\rho$, similar to the compositional boundary condition used by  Seppecher \cite{Seppecher} and Jacqmin \cite{Jacqmin}.  Consequently the assumption of Mauri \cite{Mauri} that $\delta ({\nabla}\rho \cdot \vec n)$ is independent from $\delta\rho$  is contradicted by the his conclusion that $({\nabla}\rho \cdot \vec n)$ is a function of $\rho$. 


\section{Quasi-incompressible mixture of partially miscible liquids}
We now consider a quasi-incompressible binary mixture of liquids \citep{Lowengrub}. The mixture has two equilibrium concentrations $c_\alpha$ and $c_\beta$  ($c$ is the mass fraction of one of the components) corresponding to liquid-liquid equilibrium of  two uniform phases.  The definition of a quasi-incompressible mixture is that the density $\rho(c)$ of the mixture is a function of the mass fraction $c$, but is independent of the pressure. The wall free energy density $\bar{f}_w(c)$ is function of $c$ as well. While the bulk specific free energy is a function of $c$ and gradients of $c$ viz. $\bar{f}=\bar{f}(c,{\nabla}c,{\nabla}{\nabla}c\dots)$. The free energy of the system $F=\int_V \rho \bar{f}d^3x +\int_{\partial V}\bar{f}_wd^2x$ should now be minimized with the constraint of total mass conservation ($\int_V\rho d^3x=$ constant) and mass conservation for one of the two components ($\int_V c\rho d^3x=$ constant). We furthermore assume a Taylor expansion similar to  equation (\ref{Taylor}) and follow the same procedure as for a single component. Results concerning the chemical potential and reversible stress tensor are identical to the results given by \cite{Lowengrub}. For the natural compositional boundary condition we now find the differential equation for the normal gradient $(\nabla c\cdot \vec n)$ of composition at the surface $\partial V$:
\begin{eqnarray}\label{eq:Liquid_Liquid}
\frac{  d\bar{f}_w}{dc}&=&\left\{\left[\frac{d}{dc}\left(\rho \left(\frac{\partial\bar{f}}{\partial{\nabla}^2c} \right)_0\right)\right]-\rho\left(\frac{\partial\bar{f}}{\partial |{\nabla} c|^2}\right)_0\right\}({\nabla} c \cdot \vec n)\\ \nonumber
&-&\rho\left(\frac{\partial\bar{f}}{\partial{\nabla}^2c}\right)_0\frac{d({\nabla} c \cdot \vec n)}{dc}.
\end{eqnarray}
An initial condition for integration of this  differential equation could be obtained by considering the limit of a fully saturated wall (with one of the components). This corresponds to the mass fraction $c=1$ for $d\bar{f}_w/dc<0$. For the limit $c=1$ it seems rather logical to assume that  $({\nabla} c \cdot \vec n)=0$ because  the mass fraction cannot become larger than one ($c\leq 1$).  As in the single component case the integration is carried out in the thermodynamic variable space. The actual range  $c_\alpha\leq c\leq c_\beta$ in which $c$ varies in a physical system will in general not include $c=1$. This model could be applied for convenience. An example of the integration in thermodynamic variable space is provided in the next section.

In the case of a compressible binary mixture we have $\bar{f}=\bar{f}(\rho, {\nabla}\rho, c,{\nabla}c\dots)$. Following a similar procedure one obtains a set of partial differential equations for the gradients at the wall ${\nabla\rho\cdot \vec n}$ and ${\nabla} c{\cdot \vec n}$ as functions of $\rho$ and $c$.

\section{The influence of the extra surface free energy term}

%
%
%
Here we derive a model for the excess free energy $\Delta F$ of a semi infinite two phase binary mixture of partially miscible fluids in contact with a planar surface that takes into account the surface term neglected by Cahn \cite{Cahn}. For convenience in comparing the results obtained here to the original paper \cite{Cahn} we consider the free energy per unit volume $\hat f$ to b e a function of the molar fraction $c_m$ of one of the components. We do this for the particular case of an incompressible regular solution model as described in the paper of Cahn and Hilliard \cite{Cahn_Hilliard}. This will allow us to compare the term neglected with other terms in the model giving us some insight into whether or not this term is justifiably neglected by  Cahn \cite{Cahn}. 

We define the excess free energy $\Delta F\equiv F-F_0$ per unit area as the free energy $F$ per unit area of a semi infinite two phase binary mixture of partially miscible fluids in contact with a planar surface  minus that of a semi infinite uniform reference system $F_0$. Hence this is the energy needed to create the fluid in contact with the wall out of a uniform reference fluid of concentration $c_{m 0}$, the molar fraction far away from the wall. The reference state is one for which the fluid has no interaction with the wall. Thus we have 
\begin{equation}\label{eq:F}
F=\hat f_w(c_{m\,s})+\int_0^\infty \hat f dx.
\end{equation}
The wall free energy per unit surface $\hat f_w$ is a function of the molar fraction $c_{m,\ s}$ of the fluid adjacent to the wall. The free energy per unit volume $\hat f$ can be expanded in a Taylor series around the uniform state which for an isotropic fluid yields \cite{Cahn_Hilliard}
\begin{equation}\label{eq:Taylor}
\hat f=\hat f_0(c_m)+\kappa_1\left(\frac{d^2 c_m}{d x^2}\right)+\kappa_2\left(\frac{d c_m}{d x}\right)^2.
\end{equation}
Where $\kappa_1\equiv\left(\partial \hat f/\partial d^2 c_m / d x^2\right)_0$ and $\kappa_2\equiv\left(\frac{\partial^2 \hat f}{\partial (d c_m / d x)^2}\right)_0/2$, the subscript $0$ specifies functions of $c_m$ (viz. not of $d^2 c_m / d x^2$ and $\left(d c_m/d x\right)^2$). Substituting equation (\ref{eq:Taylor}) into equation (\ref{eq:F}) yields
\begin{equation}\label{eq:FT}
F=\hat f_w(c_{m\,s})+\int_0^\infty \left[\hat f_0(c_m)+\kappa_1\left(\frac{d^2 c_m}{d x^2}\right)+\kappa_2\left(\frac{d c_m}{d x}\right)^2\right] dx.
\end{equation}
Performing partial integration on the second term in the integrant and using $\lim_{x\rightarrow\infty}\frac{d c_m}{d x}\rightarrow0$ and $\lim_{x\rightarrow0}\frac{d c_m}{d x}\rightarrow\left(\frac{d c_m}{d x}\right)_{c_{m\,s}}$ yields
\begin{equation}
F=\hat f_w(c_{m\,s})-\kappa_1\left(\frac{d c_m}{d x}\right)_{c_{m\,s}}+\int_0^\infty \left[\hat f_0(c_m)+\kappa\left(\frac{d c_m}{d x}\right)^2\right].
\end{equation}
Where $\kappa\equiv\kappa_2-d\kappa_1/dc_m$. For the particular case considered i.e. the regular solution model of Cahn and Hilliard \cite{Cahn_Hilliard} we have $\kappa_1=-\kappa c_m$ where $\kappa$ is a positive constant and $\kappa_2=0$.
The free energy of the reference state is
\begin{equation}\label{eq:F0}
F_0=\int_0^\infty \hat f_0(c_{m\,0}) dx,
\end{equation}
where $c_{m\,0}\equiv\lim_{x\rightarrow\infty} c_m $. Thus defining $\Delta f\equiv \hat f_0(c_m)-\hat f_0(c_{m\,0})$ the excess free energy $\Delta F$ can be written as:
\begin{equation}
\Delta F=\hat f_w(c_{m\,s})-\kappa_1\left(\frac{d c_m}{d x}\right)_{c_{m\,s}}+\int_0^\infty \left[\Delta \hat f+\kappa\left(\frac{d c_m}{d x}\right)^2\right].
\end{equation}
with $\kappa_1=-\kappa c_m$. Taking the variation of $\Delta F$ we have
\begin{eqnarray}
\delta \Delta F&=&\left\{\frac{d \hat f_w}{d c_{m\,s}}-\frac{d}{d c_{m\,s}}\left[\kappa_1\left(\frac{d c_m}{d x}\right)_{c_{m\,s}}\right]-2\kappa\left(\frac{d c_m}{d x}\right)_{c_{m\,s}}\right\}\delta c_{m\,s}\nonumber\\
&+&\int_0^\infty \left[\frac{d \Delta \hat f}{d c_m}-2\kappa\frac{d^2 c_m}{d x^2}\right]\delta c_m dx\label{eq:deltaF}.
\end{eqnarray}
In equilibrium we have $\delta \Delta F=0$ where surface term of the variation and the bulk term should vanish independently \cite{CourantH}. Setting $\delta c_{m\,s}=0$ and $\delta c_m\neq0$ we find the equilibrium condition for the bulk
\begin{equation}
\frac{d \Delta \hat f}{d c_m}=2\kappa\frac{d^2 c_m}{d x^2}.
\end{equation}
We can integrate this ordinary differential equation using the boundary condition $\Delta \hat f=0$ and $d c_m / d x=0$ in the limit $x\rightarrow\infty$ to find:
\begin{equation}
\Delta \hat f=\kappa\left(\frac{d c_m}{d x}\right)^2.
\end{equation}
Which can be rewritten to find 
\begin{equation}\label{eq:Bulk}
\frac{d c_m}{d x}=-\sqrt{\frac{\Delta \hat f}{\kappa}}
\end{equation}
where we have chosen the negative root because we take $c_{m\,s}>c_m>c_{m\,0}$ in this model (the wall is partially wetting) n.b. $c\in\Omega\equiv\{c\,|\,0\leq c\leq 1\}$.

The surface tension $\gamma$ is the minimum in excess free energy of the surface per unit area  \cite{Cahn} viz.
\begin{equation}
\gamma=\hat f_w(c_{m\,s})-\kappa_1\left(\frac{d c_m}{d x}\right)_{c_{m\,s}}+\int_{c_{m\,0}}^{c_{m\,s}} 2\sqrt{\kappa\Delta \hat f} dc_m
\end{equation}
Which except for the extra surface term $-\kappa_1\left(d c_m/d x\right)_{c_{m\,s}}$ corresponds to equation (10a) of Cahn \cite{Cahn}. Using $\hat f_w(c_{m\,s})=\hat f_w(c_{m\,0})+\int_{c_{m\,0}}^{c_{m\,s}}\left(d \hat f_w / d c_{m\,s}\right) dc_m$ we can rewrite this equation as follows
\begin{equation}\label{eq:gamma}
\gamma=\hat f_w(c_0)-\kappa_1\left(\frac{d c_m}{d x}\right)_{c_{m\,s}} +\int_{c_{m\,0}}^{c_{m\,s}}\left[\frac{d \hat f_w}{d c_{m\,s}}+ 2\sqrt{\kappa\Delta \hat f} \right]dc_m.
\end{equation}
This equation corresponds to equation (10b) of Cahn \cite{Cahn} with an added surface free energy term $-\kappa_1\left(d c_m/d x\right)_{c_{m\,s}}$.
Setting $\delta c_m=0$ and $\delta c_{m\,s}\neq0$ in equation (\ref{eq:deltaF}) yields
\begin{equation}\label{eq:NB1}
\frac{d}{d c_{m\,s}}\left(\kappa_1\left(\frac{d c_m}{d x}\right)_{c_{m\,s}}\right)+2\kappa\left(\frac{d c_m}{d x}\right)_{c_{m\,s}}=\frac{d \hat f_w}{d c_{m\,s}}
\end{equation}
the so called natural boundary condition at the wall.

We assume $f'_w<0$ i.e. the surface is partially wetting. For simplicity we assume $\hat f_w'=\text{constant}$. The regular solution model of Cahn-Hilliard \cite{Cahn_Hilliard} results in $\kappa_1=-\kappa c_m\,\,\forall c_m\in\Omega$ and $\kappa_2=0$. Substituting this into equation (\ref{eq:NB1}) and defining $\hat f_w'\equiv d \hat f_w / d c_{m\,s}$, we have
\begin{equation}\label{eq:NB2}
\frac{d}{ d c_{m\,s}}\left(\frac{d c_m}{d x}\right)_{c_{m\,s}}-\frac{1}{c_{m\,s}}\left(\frac{d c_m}{d x}\right)_{c_{m\,s}}=-\frac{\hat f_w'}{\kappa c_{m\,s}}.
\end{equation} 
Which is a first order ordinary differential equation. Solving this equation and assuming $d c / d x=0$ for $c_{m\,s}=1$ (as proposed for the quasi-incompressible case), yields
\begin{equation}\label{eq:fwprimesolution}
\left(\frac{d c_m}{d x}\right)_{c_{m\,s}}=\frac{\hat f_w'}{\kappa}\left(1-c_{m\,s}\right).
\end{equation}
We will use this to investigate the importance of the surface term $-\kappa_1\left(d c_m/d x\right)_{c_{m\,s}}$  in equation (\ref{eq:gamma}). That is we have
\begin{equation}
\gamma=\hat f_w(c_{m\,0})+\kappa c_{m\,s}\left(\frac{d c_m}{d x}\right)_{c_{m\,s}}+\left(c_{m\,s}-c_{m\,0}\right)\frac{d \hat f_w}{d c_{m\,s}} +\int_{c_{m\,0}}^{c_{m\,s}} 2\sqrt{\kappa\Delta \hat f} dc_m\label{eq:gammacomp}.
\end{equation}
where for convenience we have used the assumption that $d \hat f_w / d c_{m\,s}$ is a constant. Substituting equation (\ref{eq:fwprimesolution}) into equation (\ref{eq:gammacomp}) yields
\begin{equation}\label{eq:gammacomp1}
\gamma=\hat f_w(c_{m\,0})+c_{m\,s}\left(1-c_{m\,s}\right)\frac{d \hat f_w}{d c_{m\,s}}+\left(c_{m\,s}-c_{m\,0}\right)\frac{d \hat f_w}{d c_{m\,s}} +\int_{c_{m\,0}}^{c_{m\,s}} 2\sqrt{\kappa\Delta \hat f} dc_m.
\end{equation}
The second term on the right hand side of equation (\ref{eq:gammacomp1}) is a extra surface free energy term omitted by Cahn \cite{Cahn}. 

We can now go compare the term neglected by Cahn \cite{Cahn} to the third term on the right hand side of equation (\ref{eq:gammacomp1}), which is a bulk term also present in Cahn's expression \cite{Cahn}, taking the fraction of the extra surface term with respect to the bulk term yields
\begin{equation}
\Lambda\equiv\frac{c_{m\,s}(1-c_{m\,s})}{(c_{m\,s}-c_{m\,0})}=\frac{1-c_{m\,s}}{1-\frac{c_{m\,0}}{c_{m\,s}}}.
\end{equation}
Which in general will not be small for $c_{m\,s}\in(0,1)$ and $c_{m\,0}<c_{m\,s}$. We conclude that one cannot make the a priori assumption that the results obtained by Cahn and Hilliard \cite{Cahn_Hilliard} for cases where boundary effects are neglected can be extended to problems that involve surface effects e.g. critical point wetting \cite{Cahn}. 

\section{Discussion and conclusion}
The square gradient model always implies a partial integration of the Laplacian term in the van der Waals expansion of the free energy taking non-local effects into account. This generates a surface term that is neglected in the literature without justification. For binary regular solutions  \cite{Cahn_Hilliard} and \cite{Verschueren}  show that the Laplacian term ${\nabla}^2c$ term is the leading order term in the van der Waals expansion. Hence one cannot use the argument that the term in ${\nabla}^2c$ (or  ${\nabla}^2\rho$) is negligible compared to the square gradient term in the bulk. Therefore one cannot avoid the partial integration of the Laplacian term as a step in the diffuse interface theory. When considering contact line motion or wetting phenomena one should therefore quantify the surface contribution to the wall free energy resulting from this partial integration of the Laplacian term. 

We have obtained a compositional natural compositional boundary condition at the wall for the Cahn-Hilliard-van der Waals diffuse interface model in the case of a single component vapor-liquid  system and for a two phase system in a quasi-incompressible binary mixture of partially miscible fluids. Specifically we did not neglect the surface term proportional to $(\partial \bar{f}/\partial{\nabla}^2c)_0$ in the free energy as done by Cahn and Hilliard (\citep{Cahn_Hilliard}, \citep{Cahn}).  This compositional boundary condition yields a differential equation for the wall normal gradient $({\nabla}c{\cdot \vec n})$ (or $({\nabla} \rho{\cdot \vec n})$) that should be integrated in the thermodynamic space. Determining a physically relevant initial condition for the integration of this differential equation  (\ref{vapor_liquid}) or (\ref{eq:Liquid_Liquid}) will depend on the thermodynamic model used for the bulk of the fluid and for the fluid-wall interaction. We proposed a possible initial condition.

At the present time the use of a near critical equation of state (Ginsburg-Landau \citep{Jacqmin}) has become common practice. In order to avoid critical point wetting \citep{Cahn} a particular expression for $\bar f_w$ proposed by \cite{Jacqmin} is used, imposing $({\nabla}c{\cdot \vec n})=0$ for $c=c_\alpha$ and $c=c_\beta$. These assumptions considerably simplify the numerical solution of the problem but make it difficult to use the theory for quantitative prediction of the behavior of real physical systems. Improving the natural compositional boundary condition for such simplified models has probably little added value.   However for more realistic physical models this could be significant and deserves to be investigated.


\bibliography{Lio_Diffuse_BC_POF}

\end{document}